\documentclass[aps,prd,preprint,notitlepage,nofootinbib,superscriptaddress]{revtex4-2}
\usepackage[T1]{fontenc}
\usepackage{amsmath,amssymb,mathtools,bm}
\usepackage{graphicx}
\usepackage{booktabs}
\usepackage{lmodern}
\input{glyphtounicode}
\newcommand{\ii}{\mathrm{i}}
\newcommand{\Ernst}{\mathcal{E}}
\newcommand{\Dr}{\Delta_r}
\newcommand{\Dx}{\Delta_x}
\newcommand{\Lam}{\Lambda}
\newcommand{\Kretsch}{\mathcal{K}}
\newcommand{\Ord}{\mathcal{O}}
\newcommand{\dd}{\mathrm{d}}
\newcommand{\sgn}{\operatorname{sgn}}
\newcommand{\RePart}{\operatorname{Re}}
\newcommand{\Range}{\operatorname{Range}}
\begin{document}
\title{Kerr--NUT--Levi-Civita geometries from Ernst inversion: axis structure, curvature singularities, and the Manko--Ruiz parameter}
\author{Haryanto M. Siahaan}
\email{haryanto.siahaan@unpar.ac.id}
\affiliation{Program Studi Fisika, Universitas Katolik Parahyangan,
Jalan Ciumbuleuit 94, Bandung 40141, Indonesia}

\begin{abstract}
	We construct and analyze stationary, axisymmetric vacuum metrics
	obtained by magnetic Ernst inversion of Kerr--NUT with Manko--Ruiz
	parameter $C$ and pre-inversion twist constant $\beta$.  The
	transformation lies in the Ehlers orbit; our contribution is the
	NUT-dependent geometry and the roles of $C$ and $\beta$ in its axis,
	horizon, singularity, and azimuthal-CTC structure.  The inversion
	preserves the canonical Weyl radius, the signed WLP numerator $F$, and
	the sign of $g_{\phi\phi}$ on regular domains.  At the selected pole
	$x=\sigma$, $C=-\sigma$, the local axis condition and conicity are
	controlled by $\chi_\sigma^{(\beta)}=\beta-2m(2\sigma a+3l)$.  Exterior
	zeros of the chosen seed Ernst representative obey the exact criterion
	$-\beta\in\mathcal R_C$; for the sampled families numerical traces
	yield half-line ranges with closed-form corner endpoints.  For
	$D=a^2+l^2-2alC\ne0$, $\Lambda_\beta\Sigma$ has a finite nonzero
	seed-ring limit.  High-precision calculations find direction-independent
	finite limits of both quadratic Weyl invariants along the sampled rays,
	without establishing $C^2$-extendibility.  At $\beta=0$, exterior
	simple Ernst zeros are found for the sampled $C=-1,0$ cases but not for
	$C=+1$, consistently with the computed ranges.  Near one zero the
	Kretschmann scalar has a generic sixth-order blow-up; a numerical
	angular scan identifies exceptional directions of lower order.  All
	sampled points in the regular ($g_{\phi\phi}>0$) exterior are Petrov
	type~I.  Candidate horizon locations remain those of Kerr--NUT, and
	the asymptotics are Levi--Civita type.
\end{abstract}
\maketitle
\section{Introduction}
\label{sec:introduction}
The stationary, axisymmetric vacuum Einstein equations possess a nonlinear
sigma-model structure that is efficiently encoded in the complex Ernst
potential~\cite{Ernst:1968a,Ernst:1968b,Stephani:2003,GriffithsPodolsky:2009}.
Its solution-generating symmetries, including the finite-dimensional Ehlers and
Harrison subgroups~\cite{Geroch:1971,Geroch:1972,Harrison:1968,HKX:1979},
provide a systematic route
from familiar seed metrics to black holes and other compact objects embedded
in external gravitational or electromagnetic
backgrounds~\cite{Melvin:1964,Ernst:1976}.  In vacuum, the
continuous Ehlers transformation~\cite{Ehlers:1957} and its limiting elements
are especially important for rotating backgrounds~\cite{Astorino:2020enhanced,
Astorino:2022swirling,AstorinoBoldi:2023,Astorino:2025BRBM}.
A discrete representative of this symmetry action is the inversion
\begin{equation}
  \Ernst\longmapsto \Ernst^{-1}.
  \label{eq:intro-inversion}
\end{equation}
Applied in the magnetic Ernst frame to Kerr~\cite{Kerr:1963}, it produces the
Kerr--Levi-Civita (Kerr--LC) metric~\cite{Barrientos:2025}, a rotating analogue
of the Schwarzschild--Levi-Civita solution~\cite{Barrientos:2024buchdahl,%
Mazharimousavi:2025,Amirabi:2025,Mazharimousavi:2026charged}.  Care is needed with the global
reading of such transformed metrics: in the static case, solutions presented
as vacuum were recently shown to harbour a hidden distributional matter
source that becomes visible once the metric is recast in Weyl
form~\cite{HerdeiroNovo:2026}.  The
inversion should not be advertised as an independent generating mechanism:
it is a known symmetry operation.  In metric language it is a Buchdahl-type
transformation, and the Schwarzschild--LC solution was already understood in
this way in Ref.~\cite{Barrientos:2024buchdahl}; in the Ernst framework it
sits in the Ehlers orbit, and at the level of the local line element
Kerr--LC can be reached as a singular limit of Kerr in a swirling
background~\cite{Astorino:2026backgrounds}.  Such singular limits rescale
the azimuthal coordinate, so they do not by themselves fix the global
identifications; for Levi-Civita-type metrics these identifications are
physical data, and the global structures of the two constructions need not
coincide.
Vacuum metrics have likewise been produced by combining Harrison and inversion
symmetries on other seeds~\cite{Barrientos:2026inversion}.
The same group-theoretic observation applies to any seed, including the
Kerr--NUT metric studied here.
The NUT extension nevertheless raises questions that are absent from Kerr.
The NUT parameter $l$ introduces a Misner-string
structure~\cite{NUT:1963,Misner:1963,Bonnor:1969,Clement:2015}, and the
Manko--Ruiz constant $C$ controls how the string is distributed between the
two connected components of the symmetry axis~\cite{MankoRuiz:2005}.  The
choices $C=0$ and $C=\pm1$ correspond, respectively, to a symmetric placement
and to concentrating the string on one half-axis.  This is global patch data,
not an ordinary local multipole charge: locally, $C$ is associated with a
constant mixing of the stationary and axial Killing coordinates, whereas the
periodicity of $\phi$ makes the different string placements globally
inequivalent.  This bears directly on the inversion.  A shift $C\to C+\delta C$
is undone locally by the constant Killing-basis change
$t\to t-2l\,\delta C\,\phi$, under which the axial generator itself changes
exactly, $\partial_\phi\to\partial_\phi+2l\,\delta C\,\partial_t$.
Because the magnetic reduction and the inversion are defined
relative to the chosen axial Killing field, changing $C$ before inverting means
reducing along a different generator; it is not a relabeling of the
already-transformed metric.  This is why a seed constant that is locally
removable still leaves a genuine imprint on the transformed geometry.
A second subtlety is essential for the inversion.  The twist potential of a
given seed metric is defined only up to an additive real constant $\beta$.
This constant is pure gauge before the transformation, because only
$\dd\chi$ enters the seed field equations.  It is not pure gauge after
\eqref{eq:intro-inversion}: different values of $\beta$ generally produce
different transformed metrics.  That a quantity which is pure gauge for the
seed need not survive a nonlinear transformation is a general feature of the
Ehlers--Geroch action; the point here is to make it explicit for the
Kerr--NUT inverse and to follow its consequences.  Consequently, the
transformed family admits a five-parameter representation,
\begin{equation}
  (m,a,l,C,\beta),
\end{equation}
although we shall use the simple $\beta=0$ representative for the detailed
numerical analysis.  This point is decisive for any statement about the
location or existence of zeros of the seed Ernst potential.

The aim of this paper is concrete rather than a claim of a new symmetry
transformation.  We give the exact Ernst-inverted Kerr--NUT geometry and
determine, as far as possible in closed form, how $C$ and $\beta$ fix its axis
and singularity structure.  The exact results are:
(i)~the canonical Weyl radius $\rho^2$ and the signed WLP numerator
$F=e^{2\gamma}$ are unchanged by the inversion, and the sign of
$g_{\phi\phi}$ is preserved wherever the inversion is regular;
(ii)~at the selected axis ($C=-\sigma$) the entire regular-axis behaviour,
including the conical factor, is fixed by the single constant
$\chi_\sigma^{(\beta)}=\beta-2m(2\sigma a+3l)$, for arbitrary $\beta$; and
(iii)~whether the chosen seed Ernst representative has an exterior zero is
decided jointly by $C$ and $\beta$ through the condition $-\beta\in\mathcal R_C$,
with $\mathcal R_C$ a range defined in Sec.~\ref{sec:Ernst-zero}.
We add supporting numerical results:
(iv)~for generic $D=a^2+l^2-2alC\ne0$ the leading seed-ring divergence cancels
and each quadratic Weyl invariant stays finite, with a direction-independent
limit along the sampled rays, although $C^2$-extendibility is not
established; and
(v)~a simple seed Ernst zero produces, in the example studied, a generic
sixth-order curvature pole whose angular coefficient changes sign, with a
lower-order divergence along numerically identified exceptional rays.
We also give local Killing-horizon data and a Petrov analysis of sampled
regular exterior points.

All coordinates and parameters below are made dimensionless with respect to a
fixed length $L_0$.  In a convention in which the magnetic Ernst potential
has dimensions of length squared, the dimensionally restored transformation
is $\Ernst_N=L_0^4/\Ernst_0$; our formulas set $L_0=1$.
The paper is organized as follows.  Section~\ref{sec:ernst} fixes the magnetic
Weyl--Lewis--Papapetrou frame.  Section~\ref{sec:seed} gives the Kerr--NUT seed
and its general twist representative.  Section~\ref{sec:inverted} constructs
the transformed family.  Section~\ref{sec:verification} verifies the vacuum
equations and discusses useful limits.  Section~\ref{sec:geometry} analyzes
horizons, asymptotics, the seed-ring locus, Ernst-zero singularities, Petrov
type, and the axis/causal structure.  Exact seed polynomials and numerical
methods are collected in the appendices.
\section{Magnetic Ernst formulation}
\label{sec:ernst}
We use the magnetic Weyl--Lewis--Papapetrou (WLP) form\footnote{The qualifier
``magnetic'' refers to the Killing vector used in the reduction: the potential
is built on the \emph{axial} field $\partial_\phi$, so that $f=g_{\phi\phi}$,
in contrast to the ``electric'' Ernst potential built on the stationary field
$\partial_t$.  Both are purely gravitational; no electromagnetic field is
present.  The axial (magnetic) frame is the one in which the inversion
\eqref{eq:Ernst-inversion} produces Levi-Civita asymptotics.} adapted to the
axial Killing field $\partial_\phi$,
\begin{equation}
\begin{split}
  \dd s^2={}&f(\dd\phi-\omega\,\dd t)^2
  -\frac{\rho^2}{f}\,\dd t^2
  +\frac{e^{2\gamma}}{f}
  \left(\frac{\dd r^2}{\Dr}+\frac{\dd x^2}{\Dx}\right),
  \\
  &\Dx=1-x^2,
\end{split}
\label{eq:wlp}
\end{equation}
where $f=g_{\phi\phi}$ and all potentials depend only on $(r,x)$.  Here $r$ is
a radial coordinate and $x=\cos\theta\in[-1,1]$ is the polar variable, so that
$x=\pm1$ are the two halves of the symmetry axis.  For the family considered
below,
\begin{equation}
  \rho^2=\Dr\Dx.
  \label{eq:rho}
\end{equation}
The magnetic Ernst potential is
\begin{equation}
  \Ernst=f-\ii\chi,
  \label{eq:Ernst-definition}
\end{equation}
where the twist is related to the rotation function by
\begin{equation}
  \partial_x\chi=\frac{f^2}{\Dx}\,\partial_r\omega,
  \qquad
  \partial_r\chi=-\frac{f^2}{\Dr}\,\partial_x\omega.
  \label{eq:twist-equations}
\end{equation}
The vacuum Ernst equation reads
\begin{equation}
\RePart(\Ernst)
\left[\partial_r(\Dr\partial_r\Ernst)
      +\partial_x(\Dx\partial_x\Ernst)\right]
 =\Dr(\partial_r\Ernst)^2+\Dx(\partial_x\Ernst)^2.
\label{eq:Ernst-equation}
\end{equation}
On a domain where $\Ernst\ne0$, the inversion
\begin{equation}
  \Ernst_N=\frac{1}{\Ernst}
  \label{eq:Ernst-inversion}
\end{equation}
is again a solution.  In terms of real potentials,
\begin{equation}
  f_N=\frac{f}{f^2+\chi^2},
  \qquad
  \chi_N=-\frac{\chi}{f^2+\chi^2}.
  \label{eq:inverted-real-parts}
\end{equation}
The sigma-model source that determines $\gamma$ is invariant under
\eqref{eq:Ernst-inversion}; with the same additive normalization of $\gamma$,
the numerator $e^{2\gamma}$ is therefore unchanged.  A caution on this symbol:
the metric numerator is a genuine positive exponential only where the axial
reduction is spacelike, $f>0$; elsewhere we use $e^{2\gamma}$ for the signed
algebraic function of the same name, and only the combination $e^{2\gamma}/f$,
which stays positive, enters the line element.  The inversion leaves this
numerator, not a globally real exponential, unchanged.  Our signature is
$(-,+,+,+)$.
\section{Kerr--NUT seed and twist representatives}
\label{sec:seed}
The Kerr--NUT seed~\cite{DemianskiNewman:1966,PlebanskiDemianski:1976} with
Manko--Ruiz parameter $C$ is
\begin{align}
\dd s_0^2={}&
\Sigma\left(\frac{\dd r^2}{\Dr}+\frac{\dd x^2}{\Dx}\right)
-\frac{\Dr-a^2\Dx}{\Sigma}\,\dd t^2
\nonumber\\
&+\frac{2\left[X\Dr-a\Dx(aX+\Sigma)\right]}{\Sigma}
  \,\dd t\,\dd\phi
+\frac{(aX+\Sigma)^2\Dx-X^2\Dr}{\Sigma}\,\dd\phi^2,
\label{eq:seed-metric}
\end{align}
where
\begin{equation}
\begin{split}
  \Sigma&=r^2+(ax+l)^2,\\
  \Dr&=r^2-2mr+a^2-l^2,\\
  X&=a\Dx-2l(x+C).
\end{split}
\label{eq:seed-functions}
\end{equation}
It is useful to introduce the $x$-independent quantity
\begin{equation}
  Q\equiv aX+\Sigma=r^2+a^2+l^2-2alC
  \label{eq:Q}
\end{equation}
and
\begin{equation}
  F\equiv Q^2\Dx-X^2\Dr.
  \label{eq:F}
\end{equation}
The magnetic WLP data are then
\begin{equation}
  f_0=\frac{F}{\Sigma},
  \qquad
  \omega_0=\frac{a\Dx Q-X\Dr}{F},
  \qquad
  e^{2\gamma}=F,
  \qquad
  \rho^2=\Dr\Dx.
  \label{eq:seed-WLP}
\end{equation}
Integrating \eqref{eq:twist-equations} gives
\begin{equation}
  \chi_0^{(\beta)}=\frac{\mathcal N(r,x)}{\Sigma}+\beta,
  \label{eq:seed-twist-beta}
\end{equation}
where $\mathcal N$ is displayed in Appendix~\ref{app:seed-polynomials}.  The
seed metric is independent of $\beta$.  Its Ernst representatives are
\begin{equation}
  \Ernst_0^{(\beta)}=f_0-\ii\chi_0^{(\beta)}.
  \label{eq:seed-Ernst-beta}
\end{equation}
For $a\ne0$, they can also be written in the factorized form
\begin{equation}
  \Ernst_0^{(\beta)}=
  \frac{\widetilde P_\beta(r,x)}{a^2(\ii r+ax+l)},
  \qquad
  \widetilde P_\beta=\widetilde P_0
  +a^2\beta\,[r-\ii(ax+l)],
  \label{eq:factorized-Ernst}
\end{equation}
with the polynomial $\widetilde P_0$ given explicitly in
Appendix~\ref{app:seed-polynomials}.  The unfactorized form
\eqref{eq:seed-Ernst-beta}, rather than \eqref{eq:factorized-Ernst}, should be
used when taking $a\to0$.
The imaginary translation
$\Ernst_0^{(0)}\mapsto\Ernst_0^{(0)}-\ii\beta$ changes neither the seed metric
nor the seed Ernst equation.  It does, however, change the inverse potential.
We therefore use the term \emph{seed twist gauge} only before inversion.  In
the transformed family, $\beta$ is a genuine transformation parameter.  The
choice $\beta=0$ used below is a simple representative with a finite
nonrotating $a\to0$ limit; finiteness alone does not uniquely select it from
all finite constants.
\section{The Ernst-inverted family}
\label{sec:inverted}
For each seed representative \eqref{eq:seed-Ernst-beta}, define
\begin{equation}
  \Ernst_N^{(\beta)}=\frac{1}{\Ernst_0^{(\beta)}}.
  \label{eq:inverted-Ernst-beta}
\end{equation}
Because $\Ernst_0^{(\beta)}=\Ernst_0^{(0)}-\ii\beta$, the transformed potentials
for different $\beta$ are related by a finite Ehlers (M\"obius) map on the
$\beta=0$ inverse,
\begin{equation}
  \Ernst_N^{(\beta)}=\frac{\Ernst_N^{(0)}}{1-\ii\beta\,\Ernst_N^{(0)}}.
  \label{eq:beta-mobius}
\end{equation}
Thus varying $\beta$ is an ordinary finite Ehlers action; the map is
nontrivial, so $\beta$ is not redundant.
Let
\begin{equation}
  \Lam_\beta\equiv
  \left(f_0\right)^2+
  \left(\chi_0^{(\beta)}\right)^2
  =\left|\Ernst_0^{(\beta)}\right|^2.
  \label{eq:Lambda-beta}
\end{equation}
Then
\begin{equation}
  f_N^{(\beta)}=\frac{f_0}{\Lam_\beta},
  \qquad
  \chi_N^{(\beta)}=-\frac{\chi_0^{(\beta)}}{\Lam_\beta},
  \qquad
  e_N^{2\gamma}=e^{2\gamma}=F.
  \label{eq:inverted-functions}
\end{equation}
The rotation potential is fixed, up to an additive constant $\omega_\star$,
by
\begin{equation}
\begin{split}
  \partial_r\omega_N^{(\beta)}
  &=\frac{\Dx}{(f_N^{(\beta)})^2}
    \partial_x\chi_N^{(\beta)},\\
  \partial_x\omega_N^{(\beta)}
  &=-\frac{\Dr}{(f_N^{(\beta)})^2}
    \partial_r\chi_N^{(\beta)}.
\end{split}
\label{eq:omegaN-quadrature}
\end{equation}
The two relations are compatible because of the Ernst equations, so on any
simply connected regular domain they determine a single-valued
$\omega_N^{(\beta)}$ up to $\omega_\star$; a closed rational primitive in
$(r,x)$ exists for $\beta=0$ but is too lengthy to display.
The transformed line element is \eqref{eq:wlp} with
$(f,\omega,e^{2\gamma})=(f_N^{(\beta)},\omega_N^{(\beta)},F)$.  Its nonzero
components can be organized as
\begin{equation}
\begin{split}
  g_{\phi\phi}&=f_N^{(\beta)},\\
  g_{t\phi}&=-f_N^{(\beta)}\omega_N^{(\beta)},\\
  g_{tt}&=f_N^{(\beta)}(\omega_N^{(\beta)})^2
  -\frac{\Dr\Dx}{f_N^{(\beta)}},\\
  g_{rr}&=\frac{\Lam_\beta\Sigma}{\Dr},
  \qquad
  g_{xx}=\frac{\Lam_\beta\Sigma}{\Dx}.
\end{split}
\label{eq:inverted-components}
\end{equation}
In particular,
\begin{equation}
  \det g=-\left(\Lam_\beta\Sigma\right)^2.
  \label{eq:metric-determinant}
\end{equation}
The regular domain of the transformed metric excludes zeros of
$\Ernst_0^{(\beta)}$, i.e. points where $\Lam_\beta=0$.
Unless stated otherwise, all formulas involving $\beta$ remain general, while
all numerical calculations below use
\begin{equation}
  \beta=0.
  \label{eq:beta-zero-choice}
\end{equation}
\section{Vacuum verification and limiting sectors}
\label{sec:verification}
There are two independent checks of Ricci flatness.  First,
\eqref{eq:inverted-Ernst-beta} satisfies the vacuum Ernst equation on every
regular domain because inversion is an exact symmetry of
\eqref{eq:Ernst-equation}.  Together with the unchanged signed WLP numerator
$F$, this
reconstructs a vacuum metric.  Second, direct assembly of
\eqref{eq:inverted-components} and use of \eqref{eq:omegaN-quadrature} gives
\begin{equation}
  R_{\mu\nu}=0
  \label{eq:Ricci-zero}
\end{equation}
identically after rational simplification.  Our numerical curvature checks
also give Ricci residuals below $10^{-50}$ in the high-precision tests
reported in Appendix~\ref{app:numerics}.
For the chosen representative $\beta=0$, the limit $l\to0$ removes all
$C$-dependence from $X$, $\mathcal N$, and the transformed metric, and the
result reduces to the Kerr--LC metric~\cite{Barrientos:2025}, up to the choice
of twist constant: our $\beta=0$ representative is the one that stays regular
as $a\to0$, which we take to be the normalization used for Kerr--LC.  This is a
consistency check, not a third proof of vacuum character.  For nonzero fixed
$\beta$, the same limit instead retains the pre-inversion imaginary translation
and need not coincide with the conventional Kerr--LC representative.
The unfactorized data \eqref{eq:seed-WLP} and
\eqref{eq:seed-twist-beta} also admit a finite $a\to0$ limit at $\beta=0$.
Because $l\ne0$ then gives Taub--NUT rather than a static spacetime, we refer
to this as the nonrotating limit.  A detailed analysis of the corresponding
Taub--NUT--Levi-Civita member is deferred to separate work.
\section{Geometry of the transformed spacetime}
\label{sec:geometry}
\subsection{Candidate Killing horizons and local horizon data}
\label{sec:horizons}
The canonical Weyl radius vanishes at the roots of $\Dr$,
\begin{equation}
  r_\pm=m\pm\sqrt{m^2+l^2-a^2},
  \label{eq:rplusminus}
\end{equation}
which are real when
\begin{equation}
  a^2\le m^2+l^2.
  \label{eq:horizon-condition}
\end{equation}
These roots identify candidate horizon rods.\footnote{In the Weyl $(\rho,z)$
description the symmetry axis $\rho=0$ decomposes into intervals (``rods''):
in the exterior chart the segment at fixed $r=r_+$ is the finite horizon
rod, and each half of the symmetry axis is a semi-infinite rod; $r=r_-$
belongs to a different analytic block.  A horizon rod is regular only if the
metric is well behaved along the segment and at its endpoints.}  Calling
$r=r_+$ a regular Killing horizon additionally requires that the transformed
metric be regular on the rod and at its endpoints.  In particular, we exclude
\begin{equation}
  Q_H\equiv Q(r_+)=r_+^2+a^2+l^2-2alC=0,
  \label{eq:PH}
\end{equation}
and any endpoint Ernst zero.  (We write $Q_H$ rather than $P_H$ to avoid
collision with $P(x)$, $\widetilde P_\beta$, and $\Delta_{\rm P}$.)  On the
interior of the rod, $Q_H\ne0$ gives $f_0=Q_H^2\Dx/\Sigma>0$, and hence
$\Lam_\beta>0$.
Using \eqref{eq:Q}, the induced area density is
\begin{equation}
  \left.\sqrt{g_{xx}g_{\phi\phi}}\right|_{r=r_+}=|Q_H|.
  \label{eq:horizon-area-density}
\end{equation}
For the coordinate period $\Delta\phi=2\pi$, and for arbitrary real $C$,
\begin{equation}
  A_{2\pi}=4\pi|Q_H|,
  \qquad
  |\kappa|=\frac{r_+-r_-}{2|Q_H|}
  =\frac{\sqrt{m^2+l^2-a^2}}{|Q_H|},
  \label{eq:horizon-area-kappa}
\end{equation}
with the orientation (the sign of $Q_H$) treated separately.
It is important to separate what is geometric from what is convention.  The
hypersurfaces $\Dr=0$ are unchanged by the inversion, and the area $A_{2\pi}$
at a fixed angular period is geometric; in the exterior Weyl chart $r=r_+$ is
the finite horizon rod, while $r=r_-$ belongs to another analytic block.  By
contrast, $\kappa$ depends on the normalization of $t$ (not on the angular
period), and, as we discuss in Sec.~\ref{sec:axis}, a rescaling of $\phi$ used
to impose elementary flatness on a selected axis segment also rescales the
area.  We therefore do not read $\kappa$ as a physical surface gravity.
The angular velocity
\begin{equation}
  \Omega_H=\left.\omega_N^{(\beta)}\right|_{r=r_+}
  \label{eq:OmegaH}
\end{equation}
is constant along the rod because
$\partial_x\omega_N^{(\beta)}\propto\Dr$ there.  Its numerical value depends
on the additive rotating-frame constant $\omega_\star$.  Since the spacetime
is not asymptotically flat, no preferred unit normalization of
$\partial_t$ is assumed.  We therefore describe \eqref{eq:rplusminus} only as
local Killing-horizon data and do not infer a global event horizon.
\subsection{Levi-Civita-type asymptotics}
\label{sec:asymptotics}
At fixed $x\in(-1,1)$ and large $r$,
\begin{equation}
  f_0=\Dx r^2+\Ord(1),
  \qquad
  \chi_0^{(\beta)}=-2l\mathcal A(x)r+\beta+\Ord(1),
  \label{eq:seed-asymptotics}
\end{equation}
where
\begin{equation}
  \mathcal A(x)=x^2+2Cx+1.
  \label{eq:Apoly}
\end{equation}
Thus
\begin{equation}
  \Lam_\beta=\Dx^2r^4\left[1+\Ord(r^{-2})\right],
  \qquad
  f_N^{(\beta)}=\frac{1}{\Dx r^2}
  \left[1+\Ord(r^{-2})\right].
  \label{eq:fN-asymptotics}
\end{equation}
The leading rotation term obtained from \eqref{eq:omegaN-quadrature} is
\begin{equation}
  \omega_N^{(\beta)}=2lP(x)r^2+\Ord(r),
  \qquad
  P(x)=x^3+3Cx^2+3x+C.
  \label{eq:omega-asymptotics}
\end{equation}
The finite constant $\beta$ first changes subleading terms.  Consequently,
\begin{equation}
\begin{split}
  g_{\phi\phi}&=\frac{1}{\Dx r^2}\left[1+\Ord(r^{-2})\right],\\
  g_{t\phi}&=-\frac{2lP(x)}{\Dx}+\Ord(r^{-1}),\\
  g_{tt}&=-\Dx^2r^4+\Ord(r^3),\\
  g_{rr}&=\Dx^2r^4\left[1+\Ord(r^{-1})\right],\\
  g_{xx}&=\Dx r^6\left[1+\Ord(r^{-1})\right].
\end{split}
\label{eq:metric-asymptotics}
\end{equation}
The exact canonical coordinates can be chosen as
\begin{equation}
  \rho=\sqrt{\Dr\Dx},
  \qquad
  z=(r-m)x.
  \label{eq:Weyl-coordinates}
\end{equation}
They satisfy the exact completion identity
\begin{equation}
  \dd\rho^2+\dd z^2
  =\Bigl[(r-m)^2-(m^2+l^2-a^2)x^2\Bigr]
  \left(\frac{\dd r^2}{\Dr}+\frac{\dd x^2}{\Dx}\right),
  \label{eq:rho-z-completion}
\end{equation}
so the $(r,x)$ block of the line element gives
\begin{equation}
  g_{\rho\rho}=g_{zz}
  =\frac{\Lam_\beta\Sigma}{(r-m)^2-(m^2+l^2-a^2)x^2}.
  \label{eq:grhorho}
\end{equation}
In fixed-angle sectors, $\rho\sim r\sqrt{\Dx}$, and the leading powers
$g_{tt}\sim-\rho^4$, $g_{\phi\phi}\sim\rho^{-2}$, and
$g_{\rho\rho}=g_{zz}\sim\rho^{4}$ are those of the
$\sigma=1$ Levi-Civita class~\cite{TycZofka:2026}; the Killing-mixing term
is subleading, $|g_{t\phi}|/\sqrt{|g_{tt}g_{\phi\phi}|}=\Ord(\rho^{-1})$,
which completes the comparison beyond the two diagonal Killing components.
The expansion is not
uniform as $x\to\pm1$; the separate axis analysis in Sec.~\ref{sec:axis}
controls that limit and its conicity.
\subsection{The seed ring locus}
\label{sec:ring}
The Kerr--NUT seed denominator
\begin{equation}
  \Sigma=r^2+(ax+l)^2
\end{equation}
vanishes in the physical angular interval only when $|l|\le|a|$.  The
interior case $|l|<|a|$ contains the ring-type seed singularity
\begin{equation}
  r=0,
  \qquad
  x=-\frac{l}{a}.
  \label{eq:ring-location}
\end{equation}
To examine the inverse metric, set
\begin{equation}
  r=s\cos\vartheta,
  \qquad
  ax+l=s\sin\vartheta,
  \qquad
  \Sigma=s^2,
  \label{eq:ring-rays}
\end{equation}
and define
\begin{equation}
  D\equiv a^2+l^2-2alC.
  \label{eq:D}
\end{equation}
For finite $\beta$ and generic $D\ne0$,
\begin{align}
  f_0&=\frac{2D^2}{a^2s}
  \left(m\cos\vartheta+l\sin\vartheta\right)+\Ord(1),
  \label{eq:ring-f0}\\
  \chi_0^{(\beta)}&=\frac{2D^2}{a^2s}
  \left(l\cos\vartheta-m\sin\vartheta\right)+\Ord(1).
  \label{eq:ring-chi0}
\end{align}
The finite constant $\beta$ is contained in the subleading term.  Hence
\begin{equation}
\Lam_\beta\Sigma\longrightarrow
\frac{4D^4(m^2+l^2)}{a^4}
  \label{eq:ring-LambdaSigma}
\end{equation}
independently of $\vartheta$, and
\begin{align}
  f_N^{(\beta)}&=
  \frac{a^2s}{2D^2(m^2+l^2)}
  \left(m\cos\vartheta+l\sin\vartheta\right)+\Ord(s^2),
  \label{eq:ring-fN}\\
  \chi_N^{(\beta)}&=-
  \frac{a^2s}{2D^2(m^2+l^2)}
  \left(l\cos\vartheta-m\sin\vartheta\right)+\Ord(s^2).
  \label{eq:ring-chiN}
\end{align}
Equation~\eqref{eq:ring-LambdaSigma} gives a finite nonzero determinant limit
when
\begin{equation}
  a\ne0,
  \qquad
  m^2+l^2\ne0,
  \qquad
  D\ne0.
  \label{eq:ring-generic-domain}
\end{equation}
This is an exact statement about the leading behaviour of the determinant, but
it does not by itself prove smooth extendibility.
The seed Kretschmann scalar is\footnote{The Kerr--NUT seed is Petrov type~D; in its principal null frame the
	only nonvanishing Weyl scalar is
	$\Psi_2=-(m+\ii l)\bigl(r+\ii(ax+l)\bigr)^{-3}$. As the seed is vacuum, its
	Kretschmann scalar equals the quadratic Weyl invariant
	$C_{abcd}C^{abcd}=48\,\RePart(\Psi_2^2)$, which gives
	\eqref{eq:seed-Kretschmann}; the $l\to0$ case reduces to the familiar Kerr
	value $48\,\RePart[\,m^2(r+\ii ax)^{-6}]$. The complex combination $m+\ii l$ is
	the usual gravitomagnetic ``complex mass'' of the NUT family. See
	Refs.~\cite{GriffithsPodolsky:2009,Stephani:2003}.}
\begin{equation}
  \Kretsch_0=48\RePart\left[
  \frac{(m+\ii l)^2}{\bigl(r+\ii(ax+l)\bigr)^6}\right],
  \label{eq:seed-Kretschmann}
\end{equation}
which has the standard $\Sigma^{-3}$ ring divergence.  For the transformed metric at $\beta=0$ we computed the curvature point by
point. At each sampled $(r,x)$ the covariant Riemann tensor was assembled from
the value of the metric together with its first and second derivatives at that
same point, which suffices to determine the Riemann tensor
(Appendix~\ref{app:numerics}). Taking a sequence of points that approach
the ring ($r\downarrow0$, $x=-l/a$) and extrapolating, the Kretschmann scalar
$\Kretsch$ approaches the finite limits listed in Table~\ref{tab:ring-limits}.
Two points should be kept in mind.  First, the extrapolated values span about
eleven orders of magnitude across the parameter sets in
Table~\ref{tab:ring-limits}, so the size of the limit is very sensitive to the
parameters.  Second, a finite value of one scalar is only weak evidence of
regularity.  We did, however, test direction independence directly, for every
row of Table~\ref{tab:ring-limits}.  Approaching the ring along the nine rays
\begin{equation}
  \vartheta\in\{-80^\circ,-60^\circ,-40^\circ,-20^\circ,0^\circ,
  20^\circ,40^\circ,60^\circ,80^\circ\}
  \label{eq:nine-rays}
\end{equation}
in the parametrization \eqref{eq:ring-rays}, at $s=10^{-3},\dots,10^{-6}$,
the Kretschmann scalar converges to a direction-independent limit for each
parameter set (for the largest-curvature set, $-9.57\times10^{8}$ from all
nine rays); the spread between directions falls off linearly in $s$ (about
one order of magnitude for each factor of ten in $s$).  The finite
Kretschmann limit
is therefore not an artifact of a single approach direction.  We also computed
the second quadratic Weyl invariant, the Pontryagin scalar
$\Kretsch^\ast=C_{abcd}{}^\ast C^{abcd}$ with
${}^\ast C_{abcd}=\tfrac12\epsilon_{ab}{}^{ef}C_{efcd}$, along the same rays
for all four rows; it also converges to a direction-independent finite value
in every case, listed in Table~\ref{tab:ring-limits}.  The sign of
$\Kretsch^\ast$ is orientation dependent: the tabulated values use the
orientation $\epsilon_{trx\phi}=+\sqrt{-g}$, and the opposite orientation
flips their sign.  In addition to the rays \eqref{eq:nine-rays}, we checked
the two distinguished directions $m\cos\vartheta+l\sin\vartheta=0$ and
$l\cos\vartheta-m\sin\vartheta=0$, where a leading term in
\eqref{eq:ring-fN}--\eqref{eq:ring-chiN} vanishes
($\vartheta\simeq-72.9^\circ$ and $+17.1^\circ$ for the largest-curvature
set); both invariants reach the same limits along them.  Thus \emph{both} quadratic
Weyl invariants stay finite at the ring in the sampled cases: a finite
$\Kretsch$ alone could in principle mask an unbounded pseudoscalar, and here it
does not.  (In Lorentzian signature neither invariant is sign-definite, which
is why tabulated values can have either sign.)  This still does not show that
\emph{every} curvature scalar stays finite, that the curvature remains bounded
in a parallelly propagated frame carried into the ring, or that the metric is
$C^2$-extendible across the ring.  Each is a stronger statement requiring a
separate analysis, and we do not claim any of them here.
\begin{table}[t]
\caption{Extrapolated ring limits of the two quadratic Weyl invariants as
$(r,x)\to(0,-l/a)$ for the $\beta=0$ representative.  The intercepts come from
overdetermined least-squares polynomial fits (degree $2$--$4$) of
$\Kretsch(s)$ and $\Kretsch^\ast(s)$ to high-precision evaluations at
$s=10^{-5},\dots,10^{-12}$ along $\vartheta=0$; changing the fit degree and
the fitting window moves the intercepts by less than $10^{-14}$ in relative
terms, so the quoted three figures are stable.  The raw value of $\Kretsch$
at $s=10^{-6}$ differs from the intercept by less than $4\times10^{-5}$
(relative) in every row, and by $3\times10^{-6}$ in the fourth row; for
$\Kretsch^\ast$ the corresponding differences are below $10^{-4}$.  This is
consistent with the linear approach in $s$.  The nine-ray direction checks use
$s=10^{-3},\dots,10^{-6}$.  $\Kretsch^\ast$ is evaluated with the orientation
$\epsilon_{trx\phi}=+\sqrt{-g}$; the opposite orientation flips its sign.
The column $D=a^2+l^2-2alC$ shows that the largest
magnitudes occur nearest the exceptional surface $D=0$.  The last row is the
parameter set of Figs.~\ref{fig:K-sections} and \ref{fig:K-scaled}.}
\label{tab:ring-limits}
\begin{ruledtabular}
\begin{tabular}{ccccc}
$(m,a,l,C)$ & $D$ & approach & extrap.\ $\Kretsch$ & extrap.\ $\Kretsch^\ast$ \\
\hline
$(1.3,0.7,0.4,+1)$ & $0.09$ & $x=-l/a$, $r\downarrow0$ & $-9.57\times10^{8}$ & $-5.41\times10^{8}$ \\
$(1.3,0.7,0.4,0)$  & $0.65$ & $x=-l/a$, $r\downarrow0$ & $-9.24$ & $-4.21$ \\
$(1.3,0.7,0.2,+1)$ & $0.25$ & $x=-l/a$, $r\downarrow0$ & $-7.03\times10^{4}$ & $+1.36\times10^{4}$ \\
$(1.0,0.9,0.5,-1)$ & $1.96$ & $x=-l/a$, $r\downarrow0$ & $-4.02\times10^{-3}$ & $-1.82\times10^{-3}$
\end{tabular}
\end{ruledtabular}
\end{table}
The generic expansion fails on
\begin{equation}
  D=0
  \quad\Longleftrightarrow\quad
  C=\frac{a^2+l^2}{2al}
  \qquad(al\ne0),
  \label{eq:Dzero}
\end{equation}
and the boundary $|l|=|a|$, where the ring reaches an axis pole, also requires
a separate analysis.  We make no regularity claim on these exceptional
subfamilies.
\subsection{Zeros of the seed Ernst potential}
\label{sec:Ernst-zero}
\begin{figure}[t]
  \includegraphics[width=0.92\linewidth]{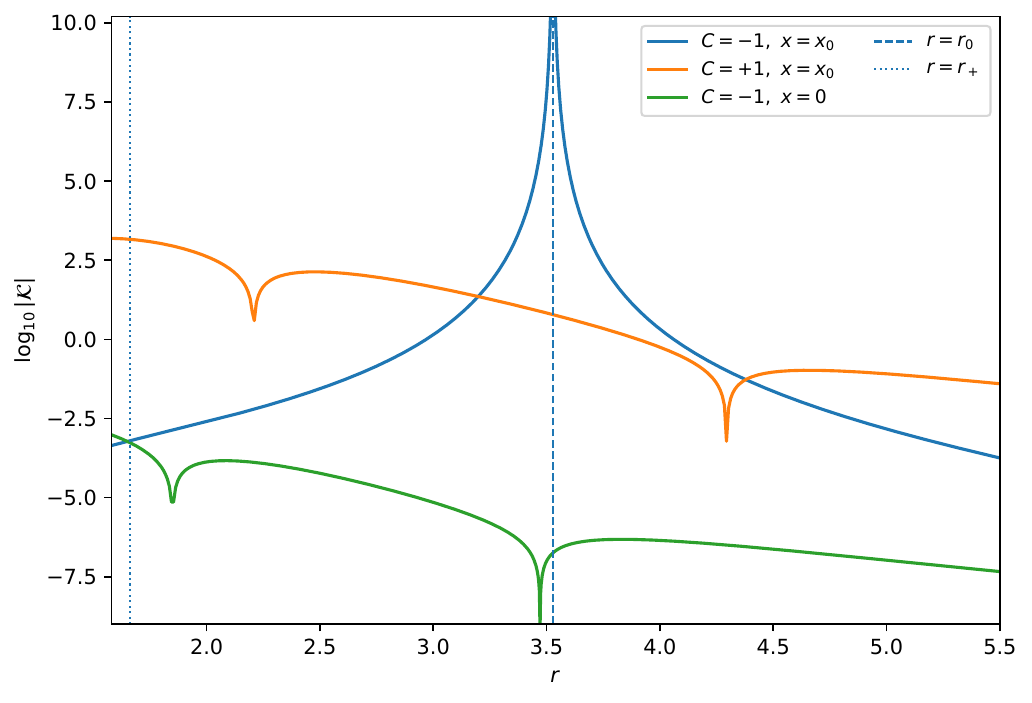}
  \caption{$\log_{10}|\Kretsch|$ for $(m,a,l)=(1,0.9,0.5)$ in the
  $\beta=0$ representative.  The $C=-1$ section through the Ernst zero
  diverges at $r=r_0$, while the displayed $C=+1$ section and the $C=-1$
  equatorial section remain finite on the plotted interval.  Zeros of
  $\Kretsch$ appear as downward cusps in the logarithmic plot.}
  \label{fig:K-sections}
\end{figure}

\begin{figure}[t]
  \includegraphics[width=0.92\linewidth]{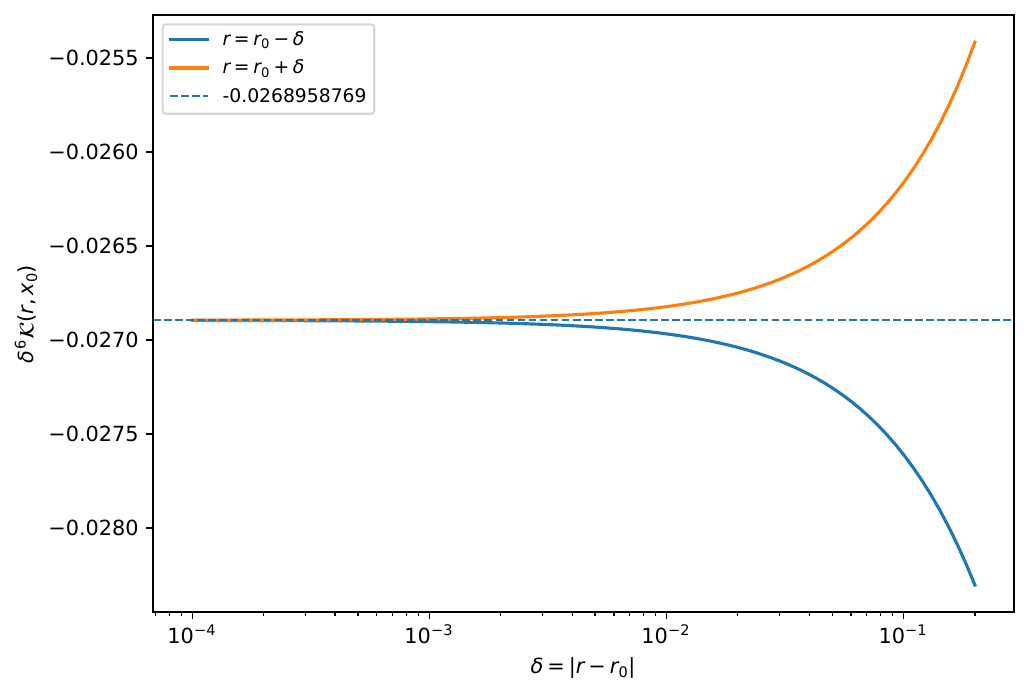}
  \caption{Scaled curvature on the two sides of the simple Ernst zero in
  Eq.~\eqref{eq:Ernst-zero-point}.  With
  $\delta=|r-r_0|$, both $\delta^6\Kretsch(r_0\pm\delta,x_0)$ approach the
  same constant, confirming the sixth-order scaling along the $x=x_0$ slice.
  The horizontal line marks the extrapolated $x=x_0$ coefficient
  $A(0)=-0.0268958769\ldots$.  As discussed in the text, the angular
  coefficient $A(\psi)$ is direction dependent, and a numerical scan
  identifies exceptional rays of lower order.}
  \label{fig:K-scaled}
\end{figure}

The inverse potential is undefined when
\begin{equation}
  \Ernst_0^{(\beta)}=0
  \quad\Longleftrightarrow\quad
  f_0=0,
  \qquad
  \frac{\mathcal N}{\Sigma}=-\beta.
  \label{eq:Ernst-zero-condition}
\end{equation}
A generic simple zero is a transverse intersection in the orbit space.  At
such a point $\Lam_\beta$ vanishes quadratically in a regular transverse
coordinate.  The transformed metric determinant vanishes there, and a local
curvature expansion generically contains $\Lam_\beta^{-3}$.  Whether the
coefficient is nonzero must be checked for the branch under consideration.
For the $\beta=0$ representative with
\begin{equation}
  (m,a,l,C)=(1,0.9,0.5,-1),
  \label{eq:representative-parameters}
\end{equation}
we find the exterior zero
\begin{equation}
  r_0=3.528319574155585\ldots,
  \qquad
  x_0=-0.938069703292712\ldots,
  \label{eq:Ernst-zero-point}
\end{equation}
while
\begin{equation}
  r_+=1.663324958071080\ldots.
\end{equation}
The Jacobian is nonzero,
\begin{equation}
  \det\frac{\partial(f_0,\chi_0^{(0)})}{\partial(r,x)}
  \bigg|_{(r_0,x_0)}
  =125.387668929\ldots,
  \label{eq:Ernst-zero-Jacobian}
\end{equation}
so the zero is simple.  Along $x=x_0$,
\begin{equation}
  \Kretsch(r,x_0)
  \sim-\frac{0.0268959\ldots}{(r-r_0)^6}.
  \label{eq:K-pole}
\end{equation}
The coefficient depends on the chosen radial coordinate and is not by itself
invariant; the geometric content is the divergence and its scaling with a
regular transverse coordinate.  Because the Jacobian
\eqref{eq:Ernst-zero-Jacobian} is nonzero, $u=f_0$ and $v=\chi_0^{(0)}$ are
valid local orbit-space coordinates, and $\Lam_0=u^2+v^2$ vanishes
quadratically at the zero.  To probe the two-dimensional structure we used
straight rays in the $(r,x)$ chart,
$(r,x)=(r_0+\delta\cos\psi,\,x_0+\delta\sin\psi)$, along which
\begin{equation}
  \Kretsch=A(\psi)\,\delta^{-6}+\Ord(\delta^{-5}),
  \label{eq:angular-coefficient}
\end{equation}
with $A(\psi+\pi)=A(\psi)$, since $\psi$ and $\psi+\pi$ are the two sides of
the same straight line and the two one-sided limits agree.  The angular
coefficient is strongly direction dependent and is \emph{not} largest on the
$x=x_0$ slice: high-precision evaluations give
$A(0^\circ)=-0.0268959\ldots$, $A(1^\circ)=+0.0063381\ldots$,
$A(90^\circ)=+2.2051\times10^{-7}$, and $A(179^\circ)=-0.0445614\ldots$.
The numerical scan finds sign-changing zero sectors of $A(\psi)$: on a
$1^\circ$ grid the sign changes in six sectors mod $\pi$.  Refining the
first of them gives the exceptional ray
\begin{equation}
  \psi_\star=0.8098478032\ldots^\circ,
  \qquad
  A(\psi_\star)=0,
  \label{eq:exceptional-ray}
\end{equation}
along which the sixth-order term cancels and the divergence drops by one
order, $\delta^5\Kretsch\to0.0214700047\ldots$.  In summary, the Kretschmann
scalar has a generic sixth-order blow-up near
the simple Ernst zero, with a direction-dependent coefficient whose
numerically located zeros give lower-order divergences along exceptional
directions.
The proper
radial distance to the singularity along the $x=x_0$ slice is finite because
$g_{rr}=\Ord((r-r_0)^2)$.
Figure~\ref{fig:K-sections} uses the $\beta=0$ curvature data.
The singular section is contrasted with the same $x=x_0$ slice at $C=+1$ and
with the $C=-1$ equatorial slice.  Figure~\ref{fig:K-scaled} shows directly
that $(r-r_0)^6\Kretsch$ tends to the same finite constant from both sides.
The role of $\beta$ is best stated without calling it a gauge of the
transformed metric.  Let $\mathcal Z_C$ denote the set of \emph{all} regular
exterior points with $f_0=0$ for fixed $(m,a,l,C)$, i.e.
$\{(r,x):r>r_+,\ -1<x<1,\ f_0=0,\ \text{the seed point is regular}\}$, taken
over every connected component, and define
\begin{equation}
  \mathcal R_C=\Range\left(
  \left.\frac{\mathcal N}{\Sigma}\right|_{\mathcal Z_C}
  \right).
  \label{eq:range-RC}
\end{equation}
Then an exterior Ernst zero exists for the transformed member labelled by
$\beta$ if and only if
\begin{equation}
  -\beta\in\mathcal R_C.
  \label{eq:range-criterion}
\end{equation}
Thus $C$ changes the range, while $\beta$ selects a level through it; the
singularity is jointly controlled by both transformation data.  This is an
exact criterion; what remains numerical is only the shape of $\mathcal R_C$ for
given parameters.  The shape can be made concrete.  For $C\ne-\sigma$ the
axis half $x=\sigma$ carries $f_0<0$ by \eqref{eq:f0-axis}.  For the six
cases studied here (two families, $C=-1,0,+1$), numerical tracing shows
that the exterior component of $\{f_0=0\}$ bounding each nonselected axis
strip runs from the rod--pole corner $(r_+,\sigma)$ toward $r\to\infty$
while approaching the axis.  Along such a component
$\mathcal N/\Sigma\sim-4l(1+C\sigma)\,r$ at large $r$; on the corresponding
nonselected poles $1+C\sigma>0$, so for the studied $l>0$ families
$\mathcal N/\Sigma\to-\infty$.  The corner limit is the exact closed-form
value
\begin{equation}
  w_\sigma=\frac{\mathcal N(r_+,\sigma)}{r_+^2+(a\sigma+l)^2},
  \label{eq:RC-endpoint}
\end{equation}
with $\mathcal N$ from \eqref{eq:N-polynomial}.  The trace also finds
$\mathcal N/\Sigma$ monotone along each component, so each component
contributes the half-line $(-\infty,w_\sigma)$, and, for the traced cases,
\begin{equation}
  \mathcal R_C=\Bigl(-\infty,\ \max_{\sigma:\,C\ne-\sigma}w_\sigma\Bigr),
  \label{eq:RC-max}
\end{equation}
where the maximum runs over the poles that actually bound a nonselected
strip.  The exact ingredients here are the criterion
\eqref{eq:range-criterion}, the large-$r$ behavior, and the corner value
\eqref{eq:RC-endpoint}; the component shape and the monotonicity are
numerical observations.  Explicitly, for
$(m,a,l)=(1,0.9,0.5)$,
\begin{equation}
  \mathcal R_{-1}=(-\infty,9.3134\ldots),\ \
  \mathcal R_{0}=(-\infty,4.1150\ldots),\ \
  \mathcal R_{+1}=(-\infty,-2.2152\ldots),
  \label{eq:RC-intervals-one}
\end{equation}
and for $(m,a,l)=(1.3,0.7,0.4)$,
\begin{equation}
  \mathcal R_{-1}=(-\infty,6.2500\ldots),\ \
  \mathcal R_{0}=(-\infty,3.0995\ldots),\ \
  \mathcal R_{+1}=(-\infty,-2.4832\ldots).
  \label{eq:RC-intervals-two}
\end{equation}
In particular, $0\in\mathcal R_C$ for $C=-1$ and $C=0$, while
$0\notin\mathcal R_{+1}$, in both families.
For $\beta=0$ and $(m,a,l)=(1,0.9,0.5)$, a numerical multistart root search
over the exterior box $r_+<r\le30$, $|x|\le0.999$ gives
\begin{equation}
\begin{array}{ccl}
 C=-1:&(r_0,x_0)=(3.528319574\ldots,-0.938069703\ldots),&\text{a zero found in the search},\\
 C=0:&(r_0,x_0)=(3.485099660\ldots,-0.982967972\ldots),&\text{a zero found in the search},\\
 C=+1:& &\text{no zero found in the search}.
\end{array}
\label{eq:C-scan-one}
\end{equation}
The same pattern is found for the second tested set
$(m,a,l)=(1.3,0.7,0.4)$, with exterior zeros
$(r_0,x_0)=(4.2560\ldots,-0.9739\ldots)$ for $C=-1$ and
$(4.3187\ldots,-0.9933\ldots)$ for $C=0$ (both with $r_+=2.4662\ldots$), and
none found for $C=+1$.  This pattern matches the intervals
\eqref{eq:RC-intervals-one}--\eqref{eq:RC-intervals-two}: the level
$-\beta=0$ lies in $\mathcal R_{-1}$ and $\mathcal R_{0}$ but not in
$\mathcal R_{+1}$, so for $C=+1$ the absence of an exterior zero follows
from the computed range (given the numerically observed monotonicity), not
merely from a search that found nothing.  These observations are still
stated as representative numerical results; the search bounds the exterior
region but is not a certified proof of the number of zeros.  A global phase diagram in the
scale-reduced parameters $(a/m,\,l/m,\,C,\,\beta/m^2)$ requires certified
elimination or interval root isolation and is beyond the present paper.
\subsection{Petrov type on the regular exterior}
\label{sec:petrov}
On a regular region with
\begin{equation}
  f_N^{(\beta)}>0,
  \qquad
  \Dr>0,
  \qquad
  \Dx>0,
  \label{eq:tetrad-domain}
\end{equation}
we use the orthonormal coframe
\begin{equation}
\begin{split}
  e^0&=\sqrt{\frac{\Dr\Dx}{f_N}}\,\dd t,
  \qquad
  e^1=\sqrt{\frac{\Lam_\beta\Sigma}{\Dr}}\,\dd r,\\
  e^2&=\sqrt{\frac{\Lam_\beta\Sigma}{\Dx}}\,\dd x,
  \qquad
  e^3=\sqrt{f_N}\,(\dd\phi-\omega_N\dd t),
\end{split}
\label{eq:orthonormal-coframe}
\end{equation}
and the associated Newman--Penrose tetrad.  From the Weyl scalars we form the
curvature invariants~\cite{Petrov:2000}
\begin{equation}
\begin{split}
  I&=\Psi_0\Psi_4-4\Psi_1\Psi_3+3\Psi_2^2,\\
  J&=\det\begin{pmatrix}
  \Psi_0&\Psi_1&\Psi_2\\
  \Psi_1&\Psi_2&\Psi_3\\
  \Psi_2&\Psi_3&\Psi_4
  \end{pmatrix},
\end{split}
\label{eq:IJ}
\end{equation}
and $\Delta_{\rm P}=I^3-27J^2$; the spacetime is algebraically special where
$\Delta_{\rm P}=0$ and type~I where $\Delta_{\rm P}\ne0$.  As a check of the
code, the same implementation returns $\Delta_{\rm P}=0$ for the Kerr--NUT seed
to working precision, correctly recovering its known type~D.  For the
transformed $\beta=0$ member $(m,a,l,C)=(1,0.9,0.5,+1)$, our curvature
convention gives
\begin{equation}
\begin{split}
  I&=-4.859807874\times10^{-6}
     -1.166453717\times10^{-6}\,\ii,\\
  J&=\phantom{-}1.052618997\times10^{-9}
     -2.322960743\times10^{-9}\,\ii
\end{split}
\label{eq:IJ-numerical}
\end{equation}
 at $(r,x)=(3,1/3)$, with
\begin{equation}
  \frac{|\Delta_{\rm P}|}{|I|^3}=0.441174931\ldots.
  \label{eq:speciality-ratio-one}
\end{equation}
At $(r,x)=(4,-1/4)$ the same ratio is
$0.358354719\ldots$.  Here $\Delta_{\rm P}$ is of order unity relative to
$|I|^3$, well away from the algebraically special value, so the spacetime is
Petrov type~I on open neighborhoods of these two points.  We also evaluated
$|\Delta_{\rm P}|/|I|^3$ on a fixed grid for the parameter sets used
here: the radial nodes
\begin{equation}
  r_j=r_++0.08\times100^{\,j/10},
  \qquad j=0,\dots,10,
  \label{eq:petrov-grid-r}
\end{equation}
combined with the angular nodes $x_k=-0.9+0.15k$, $k=0,\dots,12$, restricted
to the regular-coframe domain ($143$ points per set).  Every sampled
point is type~I ($\Delta_{\rm P}\ne0$); the smallest normalized ratios are of
order $10^{-6}$.  For six of the seven $(m,a,l,C)$ combinations scanned the
grid minimum occurs at the innermost radius $r_++0.08$ with $x=-0.9$; for
$(1,0.9,0.5,-1)$ it occurs instead at $r=3.6728\ldots$, $x=-0.9$, in the
neighborhood of the exterior Ernst zero \eqref{eq:Ernst-zero-point}.  A
separate horizon-approach scan at $(1,0.9,0.5,+1)$ shows the ratio
decreasing roughly quadratically in $r-r_+$.  We
therefore claim type~I on the sampled regular exterior away from the horizon;
we make no algebraic-type claim on the horizon rod itself, and isolated or
lower-dimensional special subsets are not excluded.  A certified classification
would require an algebraic analysis of $\Delta_{\rm P}$, which we leave open.
\subsection{Axis, Misner strings, conicity, and azimuthal CTCs}
\label{sec:axis}
At the poles $x=\sigma$, $\sigma=\pm1$,
\begin{equation}
  X_\sigma=-2l(\sigma+C),
  \label{eq:X-axis}
\end{equation}
and, outside the outer root where $\Dr>0$,
\begin{equation}
  f_0\big|_{x=\sigma}
  =-\frac{4l^2(\sigma+C)^2\Dr}
  {r^2+(a\sigma+l)^2}.
  \label{eq:f0-axis}
\end{equation}
On every regular domain $\Lam_\beta>0$,
\begin{equation}
\sgn f_N^{(\beta)}=\sgn f_0.
  \label{eq:sign-preservation}
\end{equation}
An ordinary zero of $f_N$ occurs only when $f_0=0$ and
$\chi_0^{(\beta)}\ne0$; a simultaneous zero is instead an Ernst singularity.
Thus the Manko--Ruiz choice
\begin{equation}
  C=-\sigma
  \label{eq:C-regular-pole}
\end{equation}
selects $x=\sigma$ as a candidate rotational axis, provided it is not also an
Ernst zero.  The opposite pole has $f_N<0$ for $l\ne0$, so the periodic axial
Killing orbit is timelike there and in an adjacent region.
The candidate-axis expansion can be obtained exactly.  Set
$C=-\sigma$ and
\begin{equation}
  \Sigma_\sigma=r^2+(a+\sigma l)^2.
\end{equation}
Then
\begin{equation}
  f_0=\Sigma_\sigma\Dx+\Ord(\Dx^2),
  \qquad
  \chi_0^{(\beta)}=\chi_\sigma^{(\beta)}+\Ord(\Dx^2),
  \label{eq:axis-seed-expansion}
\end{equation}
where
\begin{equation}
\chi_\sigma^{(\beta)}
=\beta-2m(2\sigma a+3l).
  \label{eq:chi-sigma-beta}
\end{equation}
For $\chi_\sigma^{(\beta)}\ne0$,
\begin{equation}
\begin{split}
  f_N^{(\beta)}&=
  \frac{\Sigma_\sigma}{(\chi_\sigma^{(\beta)})^2}\Dx
  +\Ord(\Dx^2),\\
  \chi_N^{(\beta)}&=-\frac{1}{\chi_\sigma^{(\beta)}}
  +\Ord(\Dx^2).
\end{split}
\label{eq:axis-inverted-expansion}
\end{equation}
The quadrature then gives a finite one-sided limit of $\omega_N$ and
\begin{equation}
  g_{t\phi}=-f_N\omega_N=\Ord(\Dx)\longrightarrow0.
  \label{eq:gtphi-axis}
\end{equation}
The selected pole is therefore a genuine local rotation axis, up to a conical
normalization, when
\begin{equation}
  \chi_\sigma^{(\beta)}\ne0.
  \label{eq:axis-regularity-condition}
\end{equation}
If \eqref{eq:axis-regularity-condition} fails, the entire candidate segment is
an axis Ernst zero.  For the $\beta=0$ representative, the exceptional
condition is
\begin{equation}
  m=0
  \qquad\text{or}\qquad
  2\sigma a+3l=0.
  \label{eq:axis-exception-beta0}
\end{equation}
This replaces the exceptional-axis conditions obtained in other twist
representatives.
Writing $x=\sigma\cos\theta$ near the selected pole, so that
$\Sigma=\Sigma_\sigma-a(a+\sigma l)\theta^2+\Ord(\theta^4)$ and
$\Lam_\beta=(\chi_\sigma^{(\beta)})^2+\Ord(\theta^4)$, the transverse metric is
\begin{equation}
  \dd s_\perp^2=
  \left[(\chi_\sigma^{(\beta)})^2\Sigma_\sigma+\Ord(\theta^2)\right]\dd\theta^2
  +\left[\frac{\Sigma_\sigma}{(\chi_\sigma^{(\beta)})^2}
   \theta^2+\Ord(\theta^4)\right]\dd\phi^2.
  \label{eq:axis-two-metric}
\end{equation}
The $\Ord(\theta^2)$ term in the $\dd\theta^2$ coefficient (explicitly
$-(\chi_\sigma^{(\beta)})^2a(a+\sigma l)\theta^2$) does not affect the leading
conicity, which is set by the $\theta\to0$ coefficients.
For $\phi\sim\phi+2\pi$, the circumference-to-radius ratio is controlled by
\begin{equation}
\alpha_\sigma^{(\beta)}
=\frac{1}{(\chi_\sigma^{(\beta)})^2}.
  \label{eq:conicity-general}
\end{equation}
A value $\alpha_\sigma^{(\beta)}=1$ corresponds to an ordinary, regular
rotation axis, whereas $\alpha_\sigma^{(\beta)}\ne1$ is a conical singularity
(an angular deficit or excess).  At $\beta=0$,
\begin{equation}
  \alpha_\sigma^{(0)}=
  \frac{1}{4m^2(2\sigma a+3l)^2}.
  \label{eq:conicity-beta0}
\end{equation}
This is a derived quantity in the fixed angular normalization, not an
independent modulus.  Elementary flatness on the selected segment is obtained
by $\phi=(\chi_\sigma^{(\beta)})^2\varphi$ with
$\varphi\sim\varphi+2\pi$.  In that convention the horizon area is multiplied
by $(\chi_\sigma^{(\beta)})^2$ relative to $A_{2\pi}$ in
\eqref{eq:horizon-area-kappa}.
Because \eqref{eq:sign-preservation} is exact, the inversion does not remove
the seed region of closed timelike \emph{azimuthal Killing orbits}.  For
$C=\pm1$ (and $\chi_\sigma^{(\beta)}\ne0$), one pole is a non-Ernst-singular
candidate rotation axis
and is approached with $f_N\to0^+$, while the opposite string pole has $f_N<0$.  For $C=0$, both
poles carry the NUT-string/azimuthal-CTC behavior.  This sign analysis does
not classify every possible closed timelike curve in the spacetime.
Finally, on a general $\rho=0$ rod the kernel of the Killing block is described
by a rod vector involving $\partial_t$ and $\partial_\phi$.  A complete global
rod-identification analysis requires fixing the additive constant in
$\omega_N$ and the allowed periodic identifications of time.  The present
results establish the local rotation-axis condition, conicity, and sign of
$g_{\phi\phi}$, but do not claim a global removal of the Misner string by a
single coordinate patch.
\section{Conclusions}
\label{sec:conclusions}
We have constructed the Ernst-inverted Kerr--NUT family in the magnetic Weyl
frame, keeping both the Manko--Ruiz string-placement parameter $C$ and the
additive seed-twist constant $\beta$.  The latter is gauge for the seed but a
transformation parameter after inversion; this distinction is necessary for
a consistent discussion of the transformed singularity set.  The
four-parameter Kerr--NUT--Levi-Civita geometry studied numerically in this
paper is the representative $\beta=0$ of a five-parameter representation of
the transformed family.

The construction is not independent of the Ehlers orbit, and we do not present
it as a new solution-generating method.  The concrete contribution is a set of
exact statements about the inverse geometry.  The Weyl radius and the signed
WLP numerator $F$ are unchanged, the candidate horizon radii remain $r_\pm$,
and the sign of $g_{\phi\phi}$ is preserved on every regular domain.  The exact axis
expansion shows that the selected pole $C=-\sigma$ is a non-Ernst-singular local
rotation axis when $\chi_\sigma^{(\beta)}=\beta-2m(2\sigma a+3l)$ is nonzero; in
the fixed $2\pi$ angular normalization it is elementary-flat only when
$(\chi_\sigma^{(\beta)})^2=1$, and this same single constant fixes the conical
factor.  For the Ernst-zero singularities, the
exact criterion $-\beta\in\mathcal R_C$ shows that their existence is decided
jointly by $C$ and $\beta$, not by $C$ alone; for the studied families the
range endpoint is the closed-form corner value \eqref{eq:RC-endpoint}, so the
criterion carries explicit, checkable content beyond its definition.  We
regard these three results as the main output of the paper.

The numerical results are supporting evidence, and we have been careful not to
overstate them.  At the seed ring locus, the exact limit
$\Lam_\beta\Sigma\to4D^4(m^2+l^2)/a^4$ explains why the inverse metric does not
inherit the leading seed divergence for generic $D\ne0$.  High-precision
calculations find finite, direction-independent limits of both quadratic Weyl
invariants along the sampled rays, for every parameter set tested.  Even so,
the strong parameter sensitivity of those limits calls for caution, and even
the finiteness of the two quadratic Weyl invariants is insufficient to
establish bounded curvature in a parallelly propagated frame or
$C^2$-extendibility; this is why we do not promote the ring result to a
theorem of smooth extendibility.  Similarly, a simple seed Ernst zero gives,
in the example studied, a generic sixth-order Kretschmann pole outside the
candidate outer horizon, with numerically identified exceptional directions
of lower order, and the sampled exterior is Petrov type~I; neither is
claimed as a global statement.  The exceptional sectors $D=0$ and $|l|=|a|$ remain to be
analyzed separately.

The most useful next steps are therefore concrete: a certified
$(C,\beta)$ phase diagram for exterior Ernst zeros; a local extension analysis
at $\Sigma=0$ using a complete curvature or Cartan invariant set, which would
settle the ring question that our scalar data leaves open; a separate
treatment of the exceptional $D=0$ and axis-intersection sectors; an
explicit finite-Ehlers-parameter Kerr--NUT-swirling parent; and a Weyl-form
analysis of possible distributional sources for the present family, in the
spirit of Ref.~\cite{HerdeiroNovo:2026}.  These questions
would determine whether the inverse geometry is best viewed as an isolated
Levi-Civita representative or as a singular boundary of a globally better
behaved swirling family.
\appendix
\section{Seed twist polynomial and factorized Ernst numerator}
\label{app:seed-polynomials}
The numerator in \eqref{eq:seed-twist-beta} is
\begin{align}
\mathcal N={}&-2l(x^2+2Cx+1)r^3
+2mx(ax^2-3a+3lx+6Cl)r^2
\nonumber\\
&-2l\bigl(3a^2x^2-a^2+2Ca^2x^3-2alx^3+6alx+4Cal
\nonumber\\
&\hspace{2.5cm}-3l^2x^2+l^2-6Cl^2x-4C^2l^2\bigr)r
\nonumber\\
&-2m(ax+l)\bigl(a^2x^2+a^2-alx+l^2x^2-2Calx^2-4Cal
\nonumber\\
&\hspace{4.1cm}+2Cl^2x+4C^2l^2\bigr).
\label{eq:N-polynomial}
\end{align}
For $a\ne0$, write
\begin{equation}
  \widetilde P_0=p_3r^3+p_2r^2+p_1r+p_0.
\end{equation}
The coefficients are
\begin{align}
  p_3={}&\ii a^2\Dx,
  \label{eq:p3}\\
  p_2={}&a^2\left[ax\Dx-l(3x^2+4Cx+1)\right],
  \label{eq:p2}\\
  p_1={}&2a^2m\left(ax^3-3ax+3lx^2+6Clx\right)
  \nonumber\\
  &+\ii a^2\left[a^2\Dx+4alx+4Calx^2-4Cl^2x
  +3l^2\Dx-4C^2l^2\right],
  \label{eq:p1}\\
  p_0={}&a^2\Bigl[a^3x\Dx+3a^2l\Dx+al^2(x^3-5x)
  +l^3(3x^2+1)
  \nonumber\\
  &\hspace{1.5cm}+4C^2l^2(l-ax)+8Cl^2(lx-a)\Bigr]
  \nonumber\\
  &+2\ii a^2m\Bigl[a^2(x^2+1)-alx+l^2x^2
  -2Calx^2-4Cal
  \nonumber\\
  &\hspace{3.3cm}+2Cl^2x+4C^2l^2\Bigr].
  \label{eq:p0}
\end{align}
A direct polynomial division verifies
\begin{equation}
  f_0-\ii\frac{\mathcal N}{\Sigma}
  =\frac{\widetilde P_0}{a^2(\ii r+ax+l)},
\end{equation}
and adding $\beta$ changes the numerator according to
\eqref{eq:factorized-Ernst}.
\section{Numerical curvature and Petrov checks}
\label{app:numerics}
The numerical results were regenerated consistently at $\beta=0$.  To avoid
integrating an arbitrary additive constant in $\omega_N$, the curvature was
computed from the local metric jet---that is, from the value of the metric
together with its first and second derivatives at the evaluation point, which
is the only data the Riemann tensor requires.  At each point, the exact first
derivatives of $\omega_N$ were obtained from \eqref{eq:omegaN-quadrature}, and
the second derivatives by differentiating those expressions.  Since a
constant shift of $\omega_N$ is a linear redefinition of the Killing
coordinates, it was set to zero at the evaluation point.  The resulting
metric, first derivatives, and second derivatives determine the Riemann tensor
without any path integration.
Two independent implementations were used: a standard double-precision code
and an arbitrary-precision code working in $70$--$150$ digit arithmetic.
All quoted results come from the arbitrary-precision code, which gave vacuum
(Ricci) residuals below $10^{-50}$ at every reported point and reproduced
the sixth-order Ernst-zero coefficient from both sides; the double-precision
code served as an independent cross-check away from the singular sets, near
which double precision is insufficient.  The same Petrov code returns
$I^3-27J^2=0$ for the Kerr--NUT seed, providing an internal type-D
(algebraically special) check, and the closed-form identities of the text
were verified symbolically.  All numerical statements refer to the
$\beta=0$ representative and must be recomputed for any other twist
representative.
\begin{acknowledgments}
The work is supported by LPPM-UNPAR. The author used OpenAI GPT-5.6 Pro to assist with algebraic cross-checks, numerical-code drafting, and
language revision.  The author independently reviewed and verified the
resulting material and takes full responsibility for the article. 
\end{acknowledgments}


\begin{thebibliography}{99}
\bibitem{Ernst:1968a}
F.~J. Ernst,
New formulation of the axially symmetric gravitational field problem,
Phys. Rev. \textbf{167}, 1175 (1968).
\bibitem{Ernst:1968b}
F.~J. Ernst,
New formulation of the axially symmetric gravitational field problem. II,
Phys. Rev. \textbf{168}, 1415 (1968).
\bibitem{Stephani:2003}
H.~Stephani, D.~Kramer, M.~MacCallum, C.~Hoenselaers, and E.~Herlt,
\textit{Exact Solutions of Einstein's Field Equations}, 2nd ed.
(Cambridge University Press, Cambridge, 2003).
\bibitem{GriffithsPodolsky:2009}
J.~B. Griffiths and J.~Podolsk\'y,
\textit{Exact Space-Times in Einstein's General Relativity}
(Cambridge University Press, Cambridge, 2009).
\bibitem{Geroch:1971}
R.~Geroch,
A method for generating solutions of Einstein's equations,
J. Math. Phys. \textbf{12}, 918 (1971).
\bibitem{Geroch:1972}
R.~Geroch,
A method for generating new solutions of Einstein's equation. II,
J. Math. Phys. \textbf{13}, 394 (1972).
\bibitem{Harrison:1968}
B.~K. Harrison,
New solutions of the Einstein--Maxwell equations from old,
J. Math. Phys. \textbf{9}, 1744 (1968).
\bibitem{HKX:1979}
C.~Hoenselaers, W.~Kinnersley, and B.~C. Xanthopoulos,
Symmetries of the stationary Einstein--Maxwell equations. VI. Transformations
which generate asymptotically flat spacetimes with arbitrary multipole moments,
J. Math. Phys. \textbf{20}, 2530 (1979).
\bibitem{Melvin:1964}
M.~A. Melvin,
Pure magnetic and electric geons,
Phys. Lett. \textbf{8}, 65 (1964).
\bibitem{Ernst:1976}
F.~J. Ernst,
Black holes in a magnetic universe,
J. Math. Phys. \textbf{17}, 54 (1976).
\bibitem{Ehlers:1957}
J.~Ehlers,
Konstruktionen und Charakterisierungen von L\"osungen der Einsteinschen
Gravitationsfeldgleichungen,
Ph.D. thesis, Universit\"at Hamburg (1957).
\bibitem{Astorino:2020enhanced}
M.~Astorino,
Enhanced Ehlers transformation and the Majumdar--Papapetrou--NUT spacetime,
JHEP \textbf{01}, 123 (2020), arXiv:1906.08228 [gr-qc].
\bibitem{Astorino:2022swirling}
M.~Astorino, R.~Martelli, and A.~Vigan\`o,
Black holes in a swirling universe,
Phys. Rev. D \textbf{106}, 064014 (2022), arXiv:2205.13548 [gr-qc].
\bibitem{AstorinoBoldi:2023}
M.~Astorino and G.~Boldi,
Pleba\'nski--Demia\'nski goes NUTs (to remove the Misner string),
JHEP \textbf{08}, 085 (2023), arXiv:2305.03744 [gr-qc].
\bibitem{Astorino:2025BRBM}
M.~Astorino,
Black holes in the external Bertotti--Robinson--Bonnor--Melvin
electromagnetic field,
Phys. Rev. D \textbf{112}, 104077 (2025), arXiv:2508.12908 [gr-qc].
\bibitem{Kerr:1963}
R.~P. Kerr,
Gravitational field of a spinning mass as an example of algebraically special
metrics,
Phys. Rev. Lett. \textbf{11}, 237 (1963).
\bibitem{Barrientos:2025}
J.~Barrientos, A.~Cisterna, M.~Hassaine, K.~M\"uller, and K.~Pallikaris,
A new exact rotating spacetime in vacuum: The Kerr--Levi-Civita spacetime,
Phys. Lett. B \textbf{871}, 140035 (2025), arXiv:2506.07166 [gr-qc].
\bibitem{Barrientos:2024buchdahl}
J.~Barrientos, A.~Cisterna, M.~Hassaine, and J.~Oliva,
Revisiting Buchdahl transformations: New static and rotating black holes
in vacuum, double copy, and hairy extensions,
Eur. Phys. J. C \textbf{84}, 1011 (2024), arXiv:2404.12194 [gr-qc].
\bibitem{Mazharimousavi:2025}
S.~H. Mazharimousavi,
Schwarzschild--Levi-Civita black hole,
Phys. Lett. B \textbf{861}, 139234 (2025), arXiv:2403.02365 [gr-qc].
\bibitem{Amirabi:2025}
Z.~Amirabi,
On the Schwarzschild--Levi-Civita metric,
Ann. Phys. \textbf{482}, 170204 (2025).
\bibitem{Mazharimousavi:2026charged}
S.~H. Mazharimousavi,
Charged Schwarzschild--Levi-Civita black hole via the Hassan--Sen
transformation in heterotic string theory,
Fortschr. Phys. \textbf{74}, e70121 (2026).
\bibitem{HerdeiroNovo:2026}
C.~A.~R. Herdeiro and J.~P.~A. Novo,
Vacuum, ma non troppo: hidden matter distribution in symmetry-transformed
electrovacuum spacetimes,
arXiv:2605.18967 [gr-qc].
\bibitem{Astorino:2026backgrounds}
M.~Astorino,
Black holes in rotating, electromagnetic backgrounds and topological
Kerr--Newman--NUT spacetimes,
arXiv:2604.05017 [gr-qc] (2026).
\bibitem{Barrientos:2026inversion}
J.~Barrientos, A.~Cisterna, A.~D\'iaz, and K.~M\"uller,
From Bertotti--Robinson to vacuum: New exact solutions in general relativity
via Harrison and inversion symmetries,
arXiv:2602.17581 [gr-qc] (2026).
\bibitem{NUT:1963}
E.~Newman, L.~Tamburino, and T.~Unti,
Empty-space generalization of the Schwarzschild metric,
J. Math. Phys. \textbf{4}, 915 (1963).
\bibitem{Misner:1963}
C.~W. Misner,
The flatter regions of Newman, Unti, and Tamburino's generalized Schwarzschild
space,
J. Math. Phys. \textbf{4}, 924 (1963).
\bibitem{Bonnor:1969}
W.~B. Bonnor,
A new interpretation of the NUT metric in general relativity,
Proc. Cambridge Philos. Soc. \textbf{66}, 145 (1969).
\bibitem{Clement:2015}
G.~Cl\'ement, D.~Gal'tsov, and M.~Guenouche,
Rehabilitating space-times with NUTs,
Phys. Lett. B \textbf{750}, 591 (2015), arXiv:1508.07622 [hep-th].
\bibitem{MankoRuiz:2005}
V.~S. Manko and E.~Ruiz,
Physical interpretation of NUT solution,
Class. Quantum Grav. \textbf{22}, 3555 (2005), arXiv:gr-qc/0505001.
\bibitem{DemianskiNewman:1966}
M.~Demia\'nski and E.~T. Newman,
A combined Kerr--NUT solution of the Einstein field equations,
Bull. Acad. Pol. Sci. S\'er. Sci. Math. Astron. Phys. \textbf{14}, 653 (1966).
\bibitem{PlebanskiDemianski:1976}
J.~F. Pleba\'nski and M.~Demia\'nski,
Rotating, charged, and uniformly accelerating mass in general relativity,
Ann. Phys. (N.Y.) \textbf{98}, 98 (1976).
\bibitem{TycZofka:2026}
A.~Tyc and M.~\v{Z}ofka,
Geodesic structure of the cosmological Levi-Civita spacetimes,
Class. Quantum Grav. \textbf{43}, 105010 (2026), arXiv:2606.09240 [gr-qc].
\bibitem{Petrov:2000}
A.~Z. Petrov,
The classification of spaces defining gravitational fields,
Gen. Relativ. Gravit. \textbf{32}, 1665 (2000).
\end{thebibliography}
\end{document}